# RFID-BASED Prepaid Power Meter

Rozita Teymourzadeh, Mahmud Iwan, Ahmad J. A. Abueida
Faculty of Engieering, Technology and Built Environment
UCSI University, Cheras, Kuala Lumpur, Malaysia

*Abstract—* **Electrical power meter is an important component in electric energy service. In the past, many consumers have complained about reading inaccurate of the electric meter. This research presents the development of an electrical power meter equipped with RFID reader. The RFID reader reads a valid RFID card and activates the power meter so that it can supply electricity. When the credit is about low or before the electricity is auto cut off, an SMS message will be sent to the user's hand phone to alert.**

## I. INTRODUCTION

Prepaid electricity power meter may not be popular in Malaysia though it has been used in many countries. It is designed for the electric customers to have control over their electric bills. In addition, an individual can use electricity on his own bill which is useful in some situations like hotel, apartment or moving house.

This research attempts to develop a prototype of a prepaid power meter employing RFID technology [1]. In the current RFID power meters, many have used "read" types of RFID technology to activate the meter and supply electricity to the consumers [2,3]. One can only program the credit by a given code in the RFID card. The code cannot be changed. Therefore, to top up different credits in the RFID card, one should use different RFID cards instead of one. For example, if the user wants to top up RM 10, RM 20 and RM 30 on the RFID power meter, he has to use three cards with different credits hold. This scenario causes users having to buy more RFID cards for different amount of credits used.

There exists another type of RFID card where it can read and write the data. This type of RFID card requires a special RFID reader to operate. The RFID reader can be connected to computer using USB cable. When the RFID card placed on the reader, the data can be transferred to the computer or from the computer to the RFID card.

For convenient use of RFID electrical power meter, the read and writer type of RFID device is chosen. This kind of RFID device can help to reduce the number of RFID cards used for different credits top up. One card can be programmed to load for different amount of credit.

## II. DESIGN OF THE RFID-BASED PREPAID POWER METER

The diagram of designing the electrical power meter using RFID read/write reader is shown in Figures 3 and 4.

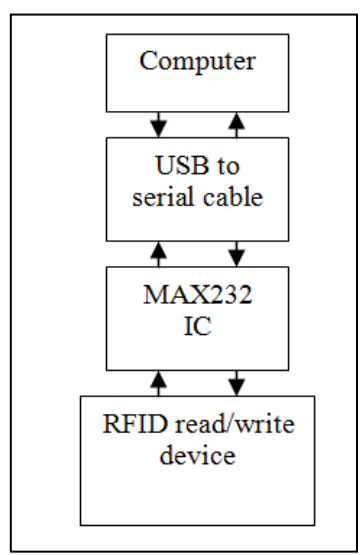

Figure 3. Connection or top up the credit in the RFID card [4].

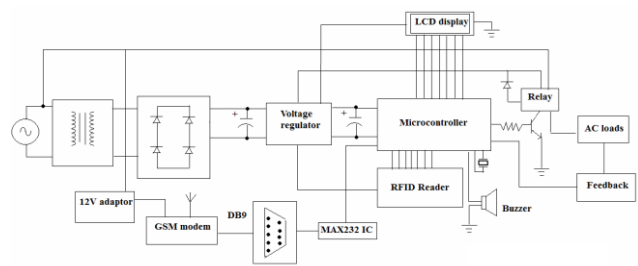

Figure 4. Basic RFID electrical power meter circuit design [2].

Figure 3 shows the connection diagram to communicate with RFID read/write device. The implementation shown in Figure 3 is mainly for top up the credit into the RFID card. Notice that in the top up process, no microcontroller is required. The communication between the device and the computer is via a direct connection to MAX232 IC. The

MAX232 regulates the voltage level between the output of the RFID read/write device and the computer. The arrows shown in Figure 3 indicate the direction flows of data. The communication between the device and the computer is in duplex or bi-direction.

Figure 4 shows the basic RFID electrical power meter designed to read the RFID card and supplies electricity to the loads.

The working principle of the RFID power meter begins from the AC 240V power source. From the block diagram, the transformer steps down the AC 240V into AC 12V. The diode connected in bridge converts 12V AC sine wave into full wave rectifier waveform. With capacitor connected at the output, the full wave rectifier waveform is converted into DC. The capacitor actually acts as filter which removes the ripple content in the DC. With voltage regulator connect across at the output of the capacitor, the DC voltage is regulated down into 5V. The 5V is then used by microcontroller, relay, MAX232 IC and LCD display.

When the RFID reader is activated or read the correct RFID card, the data from the card is transferred to the microcontroller. The microcontroller will compute the power, credit and trigger the relay so that the AC load is connected to the AC source. Both the power and credit are displayed in the LCD screen. As the credit is low, the buzzer will be activated and an SMS message will send out to the hand phone through GSM modem.

The proposed RFID power meter works very much relying on the program embedded into the microcontroller. The algorithm of C language used to detect the RFID card and activate the power meter is designed and shown in Figure 5.

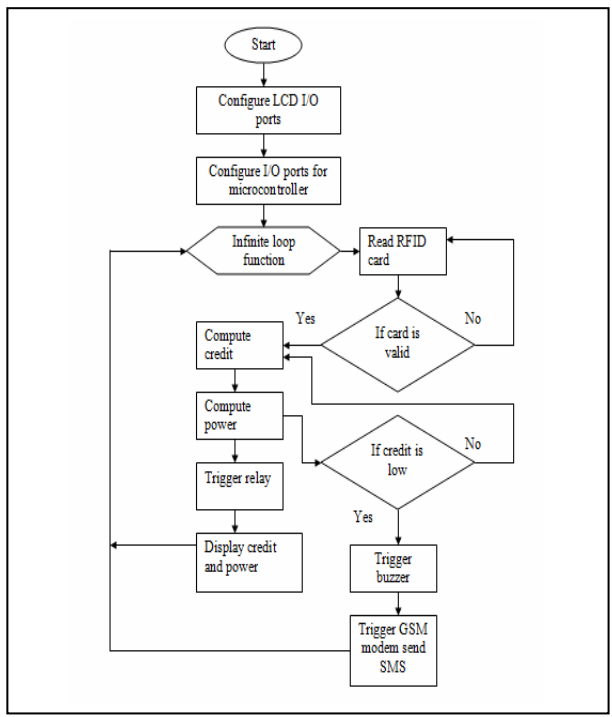

Figure 5. System algorithm design for the microcontroller work in the RFID power meter.

From the algorithm, there are two I/O ports configurations, one is for the LCD display and the other is for microcontroller. The I/O ports configuration is very important to read and write the data.

Once the microcontroller knows the I/O ports and where the data flows, the next algorithm is read the RFID card from the RFID reader. The microcontroller computes the credit deduction, displays the power and the credit, and makes decision or comparison to trigger the buzzer and the relay and GSM modem to send the SMS message. The algorithm is then repeated from the beginning when the power is cut off.

By referring to Figure 4, there is a feedback path connected at the load. This feedback is used to determine the current in the AC circuit. In other words, the feedback is actually a current sensing circuit.

For most of the digital RFID power meters, a shunt resistance method is commonly used method to determine the current. The shunt resistance will be connected in series with the load and its value is small compared to the load resistance. When the current varies, the voltage across the shunt resistor also varies. The variation of voltage gives information of the current. Figure 6 shows the shunt resistance method used as feedback system to detect the current and convert it into equivalent voltage.

From the feedback system, there is a transformer connected in shunt across the shunt resistor. The voltage drops across the shunt resistor and input to the transformer is given by [5]:

$$V = IR_{shunt} \qquad (1)$$

The capacitor connected at the output of the transformer is used to convert the AC voltage into DC voltage. This DC voltage is representing the current values and it has a unit of ampere per volt. The ampere meter "A" connected in the circuit is used for calibration and observes the output $V_{out}$ in relation to the current.

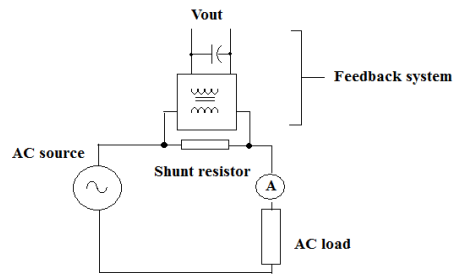

Figure 6. The feedback system to determine the current.

The calibration can be done by using different AC loads. The calibration will form a reference table for the programmer to refer to. The computation of power in the microcontroller can be written by:

$$P = 240*I \qquad (2)$$

where I refers to the table for different values of voltage. A value of 240 is a single phase AC source and P is an active power.

## III. RESULTS

Figures 7, 8, 9 and 10 show the experimental test on the RFID prepaid power meter. The circuits are constructed on the PCB and the testing was performed.

Figure 7 shows the RFID top up system, Figure 8 shows the GUI window designed to top up any values on the RFID card, Figure 9 shows the RFID power meter and Figure 10 shows the SMS received when credit is low.

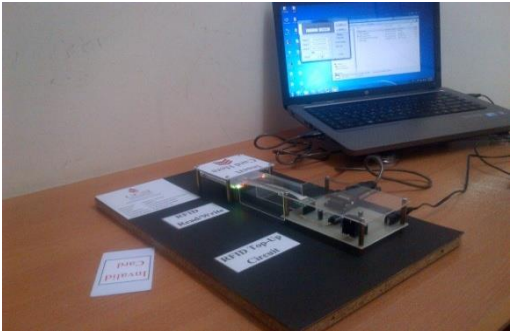

Figure 7. The RFID top up circuit

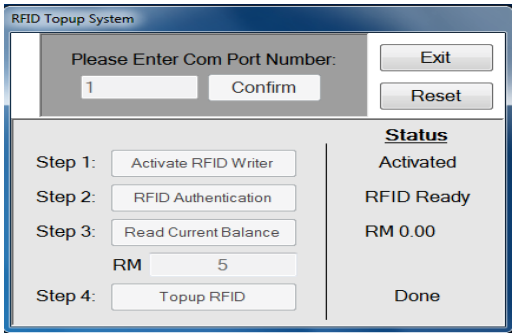

Figure 8. The top up GUI window.

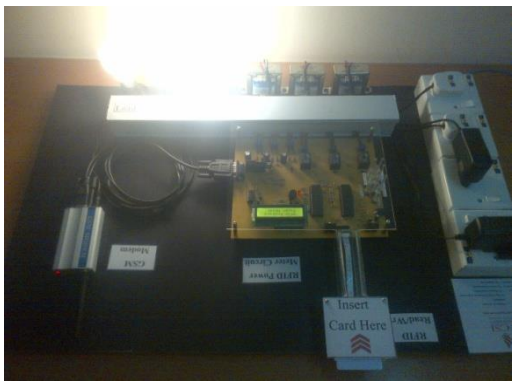

Figure 9. Test the RFID Power Meter circuit on the PCB.

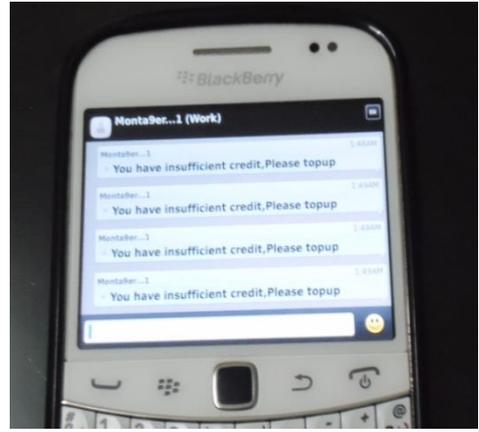

Figure 10. The SMS message received about low credit.

Table 1 shows the results of testing on the power cutoff time for the 60W, 25W and 15W light bulbs. Figure 11 shows the graphs plotted for Table I.

TABLE I. EXPERIMENTAL TEST ON LIGHT BULBS UNDER RFID POWER METER CONTROL

| Time (s) | Power consumption testing on three different light bulbs | | |
|---|---|---|---|
| | *Light bulb1* | *Light bulb 2* | *Light bulb 3* |
| 5 | 57 | 14 | 24 |
| 10 | 57 | 14 | 24 |
| 15 | 57 | 14 | 24 |
| 20 | 57 | 14 | 24 |
| 25 | 57 | 14 | 24 |
| 30 | 57 | 14 | 24 |
| 35 | 0 | 14 | 24 |
| 40 | 0 | 14 | 24 |
| 45 | 0 | 14 | 0 |
| 50 | 0 | 0 | 0 |
| 55 | 0 | 0 | 0 |
| 60 | 0 | 0 | 0 |

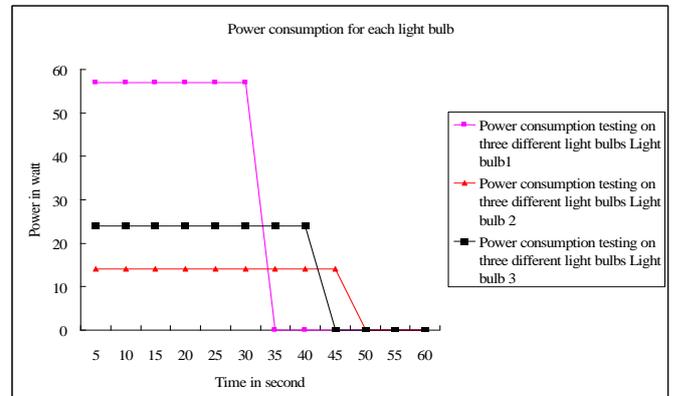

Figure 11. The operation of 60W light bulb compared to 25W, 15W light bulb

From the experimental test results, it is seen that within 1 minute, the light bulb of 60W is cutoff first. This is because the power consumption is higher compared to 25W and 15W light bulbs. The 60W light bulb takes 30 seconds to operate whereas the 25W light bulb takes 40 seconds to operate and 15W light bulb takes 45 seconds to operate. This shows that the higher the power consumed, the faster the meter will deduct the credit. In the experiment, RM 5 Ringgit is programmed for the testing.

Figure 12 shows the current sensing results for different light bulbs in the experimental test. The observation output is the DC voltage where it is important for the microcontroller to look up the current table and identify the equivalent values.

TABLE II  VOLTAGE REPRESENTING THE CURRENT FOR DIFFERENT LIGHT BULBS

| Voltage representing the current | | |
|---|---|---|
| *60W light bulb* | *25W light bulb* | *15W light bulb* |
| 3.5 | 2.2 | 1.2 |

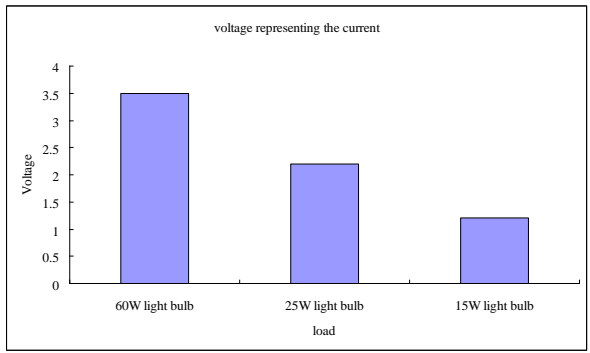

Figure 12.  The plot for Table II.

It can be seen that, the higher the load used, the more output voltage will be produced. From the graph, the connection of 60W light bulb gives 3.5V where its current is 250mA. For the 25W light bulb, the voltage produced is 2.2V and its current value is 104.2mA. The 15W light bulb gives 1.2V at the output of the current sensing circuit and its current value is 62.5mA.

## IV. CONCLUSIONS

From the experimental test, it can be seen that the operation of the RFID power meter based on the amount of credit top up is successful. This will enable the users to have control over the usage of their electric energy.

As additional feature, when the credit is about to finish, the RFID power meter will alert the users via his GSM hand phone. The electricity power supply will be cut automatically when the credit is finished.